\documentstyle[12pt,aaspp4]{article}

\def\etal{{\it et~al.\ }}

\def\refit{{\it}}

\def\Hbeta{{H$\beta$\ }}
\def\simlt{\ {\raise-.5ex\hbox{$\buildrel<\over\sim$}}\ }
\def\simgt{\ {\raise-.5ex\hbox{$\buildrel>\over\sim$}}\ }
\def\kms{{km~s$^{-1}$}}
\def\subsun{_\odot}

\begin{document} 

\title{The Circumstellar Extinction of Planetary Nebulae}

\author{Robin Ciardullo}
\affil{Department of Astronomy and Astrophysics, Penn State University,
525 Davey Lab, University Park, PA 16802}
\authoremail{rbc@astro.psu.edu}

\and

\author{George H. Jacoby}
\affil{Kitt Peak National Observatory, National Optical Astronomy
Observatories, P.O. Box 26732, Tucson, AZ 85726}
\authoremail{gjacoby@noao.edu}

\begin{abstract}
We analyze the dependence of circumstellar extinction on core mass
for the brightest planetary nebulae (PNe) in the Magellanic Clouds and M31.
We show that in all three galaxies, a statistically significant correlation 
exists between the two quantities, such that high core mass objects have 
greater extinction.  We model this behavior, and show that the relation is 
a simple consequence of the greater mass loss and faster evolution times of 
high mass stars.  The relation is important because it provides a natural
explanation for the invariance of the [O~III] $\lambda 5007$ planetary
nebula luminosity function (PNLF) with population age: bright Population~I
PNe are extinguished below the cutoff of the PNLF{}.
It also explains the counter-intuitive observation that intrinsically
luminous Population~I PNe often appear fainter than PNe from older, low-mass
progenitors.

\end{abstract}


\section{Introduction}

The modern picture of the formation and evolution of planetary nebulae (PNe)
is given by Kwok (1993, 1994).  As a star evolves up the asymptotic giant
branch (AGB), it loses mass via a slow ($\sim 15$~\kms) wind at an ever
increasing rate.  Eventually, a mass loss rate of $\dot M \approx 10^{-4}
M\subsun$~yr$^{-1}$ is achieved, and during this ``superwind'' phase,
the star becomes enshrouded by a self-generated circumstellar dust cloud.
The mass loss continues until virtually all of the star's envelope is ejected;
when this happens, the hydrogen-burning shell is exposed and the
star turns toward the blue in the HR diagram.  Shortly thereafter, the
core becomes hot enough to ionize its surroundings, and generates a 
high-speed ($v \sim 1000$~\kms) wind which compresses and shapes the newly
born planetary nebula.

A consequence of this evolutionary scenario is that the Balmer lines of 
young planetary nebulae often show the effects of extinction from dust just
outside the ionization radius.  Since PN progenitors can be anywhere from 
$M \sim 0.9 M\subsun$ (Jacoby \etal 1997) to $\sim 8 M\subsun$ (Elson \etal
1998), the amount of dust in the circumstellar region can vary greatly from 
object to object.  Moreover, because the timescale for post-AGB evolution is
strongly dependent on core mass, the extent and density of the dust 
will also be variable.  But these dependencies present us with an
opportunity: by measuring the extinction in a circumstellar 
envelope, we can learn about the properties of the progenitor star
and thereby constrain models of AGB and post-AGB evolution.

Of course, there are difficulties with this approach, particularly if the
targets are planetary nebulae in our own Galaxy.  Galactic PNe are notoriously
inhomogeneous, and have distance estimates that are extremely poor
(cf.~Pottasch 1984).  Consequently, the intrinsic properties of their
central stars cannot be accurately determined.  Similarly, in order to
measure circumstellar dust around objects in the Galaxy, one must have
a good estimate of foreground extinction.  Given the non-uniformity
of the Galaxy's cold ISM, this is, at best, a time-consuming task.
Finally, because the distances and luminosities of Galactic PNe are 
uncertain, the precise evolutionary status of many of these objects 
is controversial.  Circumstellar extinction can be expected to decrease
with time, as a consequence of nebular expansion.  Therefore, unless the
objects chosen for study are at roughly the same phase of evolution, 
the behavior of PN extinction with stellar mass cannot be studied.

The aforementioned problems prohibit us from using Galactic planetaries 
to explore the systematics of circumstellar extinction.  However, moderately
large samples of PNe with known distances and foreground reddenings 
do exist in other galaxies.  Specifically, the PNe of the Magellanic
Clouds and M31 are excellent candidates for study.  Detailed nebular 
analyses exist for $\sim 90$~extragalactic objects, and from these data, 
it is possible to obtain a large sample of circumstellar extinction
measurements and central star mass estimates.  Furthermore, since only the 
brightest PNe in these systems have been measured, most of the objects 
available for study are at roughly the same stage of evolution.  They thus 
comprise a useful database for our investigation.

Here we examine the behavior of circumstellar extinction with core mass
in sets of planetary nebulae in the Magellanic Clouds and M31.
We show that a correlation between core mass and extinction does exist,
though it is steeper than what one might predict with simple models.
We also show that this correlation provides a natural explanation for the 
invariance of the planetary nebula luminosity function (PNLF) with
population age, as the excess dust extinguishes those PNe that generate 
[O~III] $\lambda 5007$ fluxes in excess of the PNLF cutoff.  This
vitiates the conclusion of M\'endez \etal (1993) and M\'endez \& Soffner
(1997) that the PNLF cutoff must be brighter in young populations.

\section{The Extinction vs.~Core Mass Relation}
To investigate circumstellar extinction in extragalactic PNe, we restrict
ourselves to extragalactic objects whose properties have been derived via
full nebular modeling of their emission lines.  This effectively limits
our sample to two sets of data.  For the Magellanic Clouds planetaries, we 
adopt the Balmer line extinction measurements tabulated by Meatheringham \&
Dopita (1991a,b) and central star properties derived by Dopita \&
Meatheringham (1991a,b) and Dopita \etal (1997).  These papers provide us
with complete information on 73 objects: 57 in the LMC, and 16 in the SMC{}.
(We note that other samples of Magellanic Clouds planetaries do exist,
principally through the work of Kaler \& Jacoby (1990, 1991) and 
Jacoby \& Kaler (1993).  We do not include these data in our analysis, 
though their use would not affect any of our conclusions.)  For M31, we use 
the extinction estimates and central star parameters of the 15~PNe studied
by Jacoby \& Ciardullo (1999).  To insure homogeneity, all PN core masses
were re-computed by taking the central star effective temperatures and
luminosities derived from the nebular models, and re-interpolating them
onto the grid of hydrogen-burning post-AGB evolutionary tracks given by
Sch\"onberner (1983) and Bl\"ocker (1995b).  In addition, to place all
the extinction measurements on a common system, the contribution of 
Galactic dust was removed by assuming foreground reddening values of 
$E(B$$-$$V) = 0.080, 0.074$, and 0.054 for M31, the LMC, and the SMC, 
respectively (Burstein \& Heiles 1984; Caldwell \& Coulson 1985).  Here,
and throughout the paper, we assume the total extinction at H$\beta$, $c$, is 
related to differential extinction by $c = 1.47 E(B$$-$$V)$ (cf.~Kaler \&
Lutz 1985; Whitford 1958).

Plots of total logarithmic \Hbeta extinction vs.~derived core mass are
displayed in Figure~1.  From the figure, one immediately notices that there 
is considerable scatter in the relation.  This is not unexpected.  Extinction 
estimates based on measurements of the Balmer decrement may carry an 
uncertainty of up to $\sigma_c \sim 0.1$ (cf.~Ciardullo \etal 1999).  This
error, plus the error introduced by patchy extinction internal to the 
parent galaxies (Harris, Zaritsky, \& Thompson 1997; Ciardullo \etal 1988)
propagates directly into the diagrams of Figure~1 and is responsible for
the negative reddenings.  In addition, most of the PNe plotted in the figure 
are spatially unresolved, and have not been observed in the ultraviolet.  
Without the constraints provided by UV line fluxes and nebular morphology, the
models are somewhat uncertain, as are the derived positions of the central 
star in the HR diagram.  A further complication is introduced by our use of 
hydrogen-burning post-asymptotic branch evolutionary models for core mass 
determinations.  Since it is possible for PN central stars to be burning 
helium, the tracks used in this study may not be applicable for all 
objects.  Finally, even if we had the ability to place each star precisely on 
the extinction-core mass diagram, there would be intrinsic scatter due to
geometry.  Most Galactic PNe show a considerable amount of asymmetry (Balick 
1987); if the distribution of dust around the bright PNe in our extragalactic 
sample is similarly asymmetric, then orientation effects alone will broaden the
observed relation.   

Nevertheless, despite the large amount of scatter, a trend is evident
in all three galaxies.  The derived core masses of PNe with large Balmer
decrements (and therefore large \Hbeta extinctions) appear to be systematically
larger than those with small extinction values.  The relation is steep,
and there are a few high-core mass PNe in M31 and the LMC that do not obey the
rule.  But when these outlyers ($M > 0.75 M\subsun$) are omitted, the 
correlations exhibited in all three galaxies are similar.  For the LMC data,
the slope of the ordinary least squares fit bisector line is $6.3 \pm 1.3$;
for the SMC PNe, this slope is $5.6 \pm 0.7$, while for M31, the slope is
$8.5 \pm 1.6$.  Considering that the metallicities of the objects involved 
span a factor of $\sim 10$, that the ages of the parent populations are 
drastically different, and that the M31 PN observations, reductions, and 
analysis were performed completely independently of those done for the 
Magellanic Cloud PNe, the consistency of the results is striking.

\section{The Statistical Significance of the Extinction-Core Mass Correlation}

Before proceeding further, it is important to consider the statistical
significance of the relations displayed in Figure~1.  The measurement errors
on extinction and core mass are not independent.  If the extinction to a
planetary is overestimated, then its absolute emission line flux and the 
derived luminosity of its exciting central star will be overestimated.  Since
the luminosity of a young planetary nebula depends only on the mass of its
core, an error in one means an error in the other.  An artificial correlation
between extinction and core mass can then be the result.

To test the significance of the relations displayed in Figure~1, we make
the following assumptions.  First, we assume that each extinction
measurement has an uncertainty of 0.1 in $c$.  This is rather conservative.
While Ciardullo \etal (1999) have shown that extinction estimates to 
Galactic PNe have a typical error of $\sigma_c \sim 0.1$, this is often due
to the effects of atmospheric dispersion and the presence of 
non-uniformities in the distribution of dust around resolved objects.  Because 
all the PNe considered here are extragalactic, the latter problem is not an 
issue in this analysis.  Moreover, it is likely that errors due to atmospheric 
dispersion are also not a concern.  The M31 data were taken through an 
atmospheric dispersion corrector, and the Magellanic Cloud PNe were all 
observed with the slit rotated along the direction of dispersion.  An
analysis of the higher order Balmer lines in the M31 planetaries suggests
that for these objects, the true uncertainty in $c$ is $\sim 0.06$
(Jacoby \& Ciardullo 1999).  Based on external comparisons, the error
in $c$ for the Magellanic Cloud PNe is likely to be similar (Meatheringham \&
Dopita 1991a,b).   Nevertheless, because the uncertainty in $c$ is 
of critical importance for estimating the reality of the extinction-core mass
relation, we choose to be conservative and assign $\sigma_c = 0.1$.

We next assume that any error in the measurement of $c$ affects only one
quantity, the derived luminosity of the central star, and that the 
logarithmic increase (or decrease) in the star's luminosity is equal to $c$.
This is a reasonably valid approximation: although the reaction of a given 
nebular model to a change in extinction is complex, to first order, the energy
emitted at $H\beta$\ does reflect the central star's luminosity.  Furthermore,
the objects considered in this paper are all extremely bright, and
their central stars should still be moving horizontally in the HR diagram. 
Consequently, the derived mass of a PN core will depend only on the core's
luminosity, not on its effective temperature.  Uncertainties in the latter
quantity can thus be ignored.

Finally, to test the null hypothesis, we assume that extinction and central
star mass are, in reality, uncorrelated.  We randomly associate extinction
values between 0.0 and 0.8 with central star masses between $0.56 M\subsun$
and $0.66 M\subsun$, and derive the luminosities of these stars based
on the Sch\"onberner (1983) and Bl\"ocker (1995b) evolutionary tracks.  We
then place a random (Gaussian) error of $\sigma_c = 0.1$ on the extinction,
re-derive the core masses using the revised luminosity, and model the
observed PNe population of each galaxy via a series of Monte Carlo simulations.
Using 10,000 simulations per galaxy, we ask how often our model collection of 
PNe would have a Spearman rank order coefficient as large, or larger than 
that observed.  

The results for all three galaxies are similar.  Under the assumption of
a $\sigma_c = 0.1$ measurement error, the LMC correlation is significant
at the 95\% confidence level, the SMC points correlate with 97\% confidence, 
and the M31 correlation is significant at the 94\% confidence level.  
(In other words, in $\sim 95\%$ of the trials, the Spearman rank order
correlation is less than that for the real data).  If the error on 
$c$ is reduced to $\sigma_c \approx 0.06$, which, given the quality of
the data is more likely, the significance levels for the LMC and SMC data
rise to 99\%, while that for M31 increases to 96\%.  Changes in the adopted
model for the distribution of central star masses and extinctions do not
change the result significantly.  Thus, despite the scatter, the correlation
between extinction and core mass appears to be real.  It is extremely
unlikely that all three correlations are an artifact of the
analysis procedure.

\section{Modeling the Correlation}

Figure~1 confirms, at least qualitatively, the technique of measuring
PN core masses using nebular models and stellar evolutionary calculations.
The sign of the correlation is as expected: high mass PN central stars,
which presumably come from high mass progenitors, have large amounts of
circumstellar matter, and evolve quickly across the HR diagram before
their circumstellar matter has time to disperse.  Thus, they are heavily
reddened.  Low core-mass PNe have less of a circumstellar envelope and 
evolve blueward on much longer timescales.  The extinction affecting these 
objects is correspondingly less.

In addition to having the correct sign, the range of extinction displayed 
Figure~1 is also within expectations.  {\sl Hubble Space Telescope\/} images 
of bright PNe in the Magellanic Clouds shows that the median radius of these 
objects is $\sim 0.05$~pc (Dopita \etal 1996).  If we assume that the 
circumstellar envelopes obey an $1/r^2$ density law, extend to a radius of 
$\sim 0.3$~pc (Knapp, Sandell, \& Robson 1993), and have a gas-to-color-excess
ratio of $\sim 5.8 \times 10^{21}$~atoms~cm$^{-2}$~mag$^{-1}$ (Savage \& 
Mathis 1979; Knapp, Sandell, \& Robson 1993), then the amount of
extinction associated with these objects should typically be several 
tenths of a magnitude.  Again, this is what is observed.

The above calculation can be made more rigorous through the use of AGB 
and post-AGB evolutionary tracks.  To do this, we use the evolutionary
tracks of Bl\"ocker (1995a,b), which follow the mass loss and HR diagram
evolution of stars with initial masses of 3, 4, 5, and $7 M\subsun$.
First, we assume that the stellar mass loss of each model is spherically 
symmetric, and that the implied AGB wind has an expansion velocity of 
$\sim 15$~\kms.  This allows us to compute the radial profile of a star's 
circumstellar envelope at the time when it is becoming a bright 
planetary.  (For the purpose of our calculation, we assume this occurs at 
$\log T_{\rm eff} \approx 5.0$.)  Next, we adopt the canonical 
gas-to-color-excess ratio of 
$\sim 5.8 \times 10^{21}$~atoms~cm$^{-2}$~mag$^{-1}$
and assume that the ionization radius of our typical planetary nebula is 
$\sim 0.07$~pc.  This gives us the circumstellar extinction expected as a 
function of initial mass.  Finally, we use the initial-mass final-mass 
relation implied by the Bl\"ocker mass loss rates to connect the initial
stellar mass to the final core mass.  With this information, we can simulate
the data displayed in Figure~1.

The solid line drawn in the SMC panel of Figure~1 displays the results of
this calculation.  As can be seen, our simple model does produce an 
extinction-core mass relation, but with a slope that is significantly smaller 
than what is observed.  Given the amount of physics omitted from the 
computation, this is not surprising.  As the stellar core moves to the blue 
in the HR diagram, a fast wind develops which compresses the surrounding 
medium, and alters the distribution of matter close to the star.  This 
clearly is a complexity that is beyond the scope of this paper.  Moreover, 
Type~I (Population~I) planetaries in the Milky Way are often bipolar 
(Peimbert \& Torres-Peimbert 1983), hence for these objects, the assumption 
of spherical symmetry is violated.  But the most critical shortcoming of the 
model concerns the definition of the inner radius for extinction.  In our 
models, this value is well defined: extinction is only applied when light has 
passed the optical boundary of the nebula.  However, in real objects, the 
dividing line between emission and attenuation is not so discrete, and, 
since the greatest amount of extinction occurs at small radii (where the 
density of the material is highest), this over-simplification is costly.  The 
solid line of Figure~1 was computed under the assumption that the inner radius 
for extinction is 0.07~pc; this is the median nebular radius of our sample, as
as defined by the nebular models of Dopita \& Meatheringham (1991a,b) and 
Jacoby \& Ciardullo (1999).  However, as the dotted line of Figure~1
shows, if we adopt an ``effective'' radius for extinction that is 
smaller, $\sim 0.02$~pc, we can steepen the relation to match the 
observations.  (There is still a constant offset between our model and the
data, but this can be rectified in a number of ways, including adjusting 
the mass ejection rate and/or changing the gas-to-color-extinction ratio.) 
Unfortunately, without a much more detailed computation of extinction within
and around the nebulae, we cannot refine our calculation.

The qualitative agreement between the data and the computed relation
is encouraging, as it suggests that a more detailed simulation
might be able to provide new constraints on the late stages of stellar
evolution.  In particular, the amount of circumstellar extinction
surrounding a PNe is sensitive to two quantities: the total mass in the
circumstellar envelope (and therefore in the progenitor star), and the 
timescale for post-AGB evolution across the HR diagram.  If
the amount of matter in the circumstellar envelope can be measured through
extinction, it may produce an improved estimate of the initial-mass 
final-mass relation.  Similarly, if the AGB and post-AGB evolutionary 
timescales can be constrained, it may lead to a better understanding
of the planetary nebula phenomenon.  Clearly, the extinction core-mass 
relation offers some intriguing possibilities for probing stellar evolution.

\section{The Extinction-Core Mass Relation and the Planetary Nebula
Luminosity Function}
The relation displayed in Figure~1 has an interesting consequence for
the [O~III] $\lambda 5007$ planetary nebula luminosity function and
the extragalactic distance scale.  Observations of PNe in $\sim 30$ elliptical, 
spiral, and irregular galaxies have demonstrated that the PNLF is remarkably 
insensitive to stellar population (Jacoby \etal 1992; Ciardullo, Jacoby, \& 
Tonry 1993; Feldmeier, Ciardullo, \& Jacoby 1997).  Models by Jacoby (1989)
and Dopita, Jacoby, \& Vassiliadis (1992) explain why progenitor metallicity
has little effect on the PNLF, but no theory exists for the PNLF's
invariance with population age.  In fact, analyses by M\'endez \etal (1993),
Han, Podsiadlowski, \& Eggleton (1994), Jacoby (1996), and M\'endez \& 
Soffner (1997) all show that the location of the PNLF cutoff should be brighter
in young, star-forming populations.  Yet PN surveys in the LMC and three 
spiral galaxies have detected no such dependence (Jacoby, Walker, \& Ciardullo 
1990; Feldmeier, Ciardullo, \& Jacoby 1997).

Figure~1 explains the age invariance.  To first order, the strength of
a bright PN's [O~III] $\lambda 5007$ emission line depends on the amount
of energy deposited in the nebula.  Specifically, the post-AGB
evolutionary models of Sch\"onberner (1983) and Bl\"ocker (1995b) 
suggest that the maximum [O~III] $\lambda 5007$ flux attainable by a PN 
should go as roughly the square of its core mass.  Since the mass of a PN's 
core reflects the mass of its progenitor via the initial mass-final mass 
relation, this seems to imply that younger populations make brighter
planetaries.  However, as Figure~1 shows, the increased emission from
high core-mass PNe is more than made up for by the increased amount of
circumstellar extinction.   In fact, the slope of $\sim 6$ seen in the
figure guarantees that Type~I planetaries will be fainter than PNe 
derived from older stars.  This effect is not included in any
of the theoretical analyses of the PNLF, and explains why the brightest PNe
in the LMC and M101 are no brighter than those found in elliptical galaxies.  

Figure~2 demonstrates this effect quantitatively.  The dotted line in
the figure plots the predicted dependence of the PNLF cutoff with population
age, under the assumption that circumstellar extinction is not correlated
with core mass.  This calculation comes from Jacoby (1996), and assumes that 
a constant fraction (15\%) of the PN central star's luminosity is
reprocessed into [O~III] $\lambda 5007$ emission (Dopita, Jacoby, \& 
Vassiliadis 1992).  According to the model, the PNLF cutoff is a sensitive 
function of population age, with young ($\sim 1$~Gyr) stars creating 
planetaries that are $\sim 1$~mag brighter than PNe formed from old 
($\sim 10$~Gyr) stars.  It is therefore in direct conflict with observations 
of extragalactic PNe.  The solid line of Figure~2 shows the same model, except
with the effects of circumstellar extinction included via the 
extinction-core mass relation of Figure~1 (slope of $\sim 6$).  With this 
modification, the model fits the observational constraints extremely well. 
For populations older than $\sim 1$~Gyr, the PNLF is independent of age to 
within $\sim 0.1$~mag, as the increased extinction around higher mass stars 
is canceled out by the increased central star luminosity.  For younger
populations, the PNLF cutoff become fainter.  However, since real galaxies
contain a mix of stellar populations, the brightest PNe observed will 
always be those from the old and intermediate age stars.  Consequently,
distance estimates based on the PNLF will not be affected by 
Population~I planetaries.

\section{Discussion}
Aside from explaining the age invariance of the PNLF, the extinction-core
mass correlation has some other interesting consequences.  The first
involves the location of high core-mass objects on the diagram of Figure~1.
Ten objects fall well off the relation defined by the bulk of the planetary
nebulae.  These PNe are severely under-extincted for their core mass, 
and, at first glance, their presence seems to imply that the extinction
law breaks down for massive central stars.  However, an investigation of these
objects reveals that only two have derived nebular radii smaller than the
median of our sample.  According to our simple model, a larger median size
should translate into a shallower extinction law.  This is consistent with
the data.

In fact, it is relatively easy to explain the ``anomalous'' objects of 
Figure~1.  High core-mass planetaries are born on the main extinction-core 
mass relation, but their large, thick circumstellar dust cloud prevents them 
from being identified in extragalactic surveys.  Only at a later time, when 
the envelope has begun to disperse and the optical nebula has grown in size, is
the PN available for study.  By this time, the central star has completed
its trip to the blue in the HR diagram and has begun to fade.  Consequently, 
the PN will not appear ``super-luminous'' in an [O~III] $\lambda 5007$
survey.  However, because the object was extremely luminous to begin with,
it will still be bright enough to be detected photometrically, and studied 
spectroscopically.  It will therefore appear to be under-extincted in the 
extinction core-mass diagram. 

We do note that, based on the galaxies' stellar populations, we would expect 
to see a larger fraction of high core mass planetaries in the Magellanic 
Clouds than in M31.  Somewhat surprisingly, this does not seem to be the case.
Roughly 15\% of the PNe in the LMC and M31 are high core-mass objects, while
no high core-mass planetaries have been identified in the SMC{}.  The
significance of this is difficult to evaluate, since none of the PN samples 
is in any way statistically complete.  Moreover, the absence of high core-mass
PNe in the SMC may be due to random chance, since, all things being equal,
we would expect our sample to contain only $\sim 2$ of these objects.   Since
we know of no selection effect which would prevent us from seeing high core
mass PNe in the SMC, the absence of these objects is probably not significant.

Another implication of Figure~1 deals with its potential use for measuring
galaxy evolution.  Dopita \etal (1997) have shown that it is possible to 
trace the past star formation and chemical evolution of a galaxy through
measurements of its planetary nebulae.  One uses the nebular emission lines
to determine the effective temperature and luminosity of the central star.
Then, with the aid of post-AGB evolutionary tracks, one derives the PN
core mass.  Once the core mass is known, the age of the progenitor then
follows from the initial mass-final mass relation.

The key to this procedure is obtaining the mass of the PN core, and the
extinction-core mass relation is another tool for constraining this
quantity.  Because the scatter in extinction is large, age estimates
based on extinction measurements alone will probably not be possible for
individual objects.  However, the mean extinction of a population of PNe
might be a useful tool for estimating age, especially if the measurement is
done differentially with respect to the mean of other systems.
Further observations will be needed to check this possibility.

We conclude by emphasizing that in order to study the systematics of
circumstellar extinction, the sample of objects must be chosen carefully. 
The analysis in this paper was limited to only the brightest planetary nebulae.
Consequently, only newly formed objects near the peak of their UV luminosity 
were included in the samples.  With this restriction, a correlation between
extinction and core-mass can be detected.  However, as a PN evolves, its 
nebula expands, its dust envelope disperses, and its circumstellar extinction 
decreases.  Samples of PNe that span a wide range in age will therefore
not show the correlation.  As a result, PN observations in the Galaxy cannot be
used for investigations of this type.  As in other fields of study involving
planetary nebulae, the best place to investigate circumstellar extinction
is in other galaxies.

RC wishes to thank Dr.~Sidney Wolff for providing office facilities at NOAO
during part of the development of this paper.  Similarly, GHJ wishes to
thank Dr.~Peter Strittmatter for generously providing all office needs during
a sabbatical stay at the University of Arizona.  This work was supported in
part by NASA grant NAG 5-3403 and NSF grants 92-57833 and 95-29270.  

\pagebreak

\clearpage
\begin{figure}
\figurenum{1}
\plotfiddle{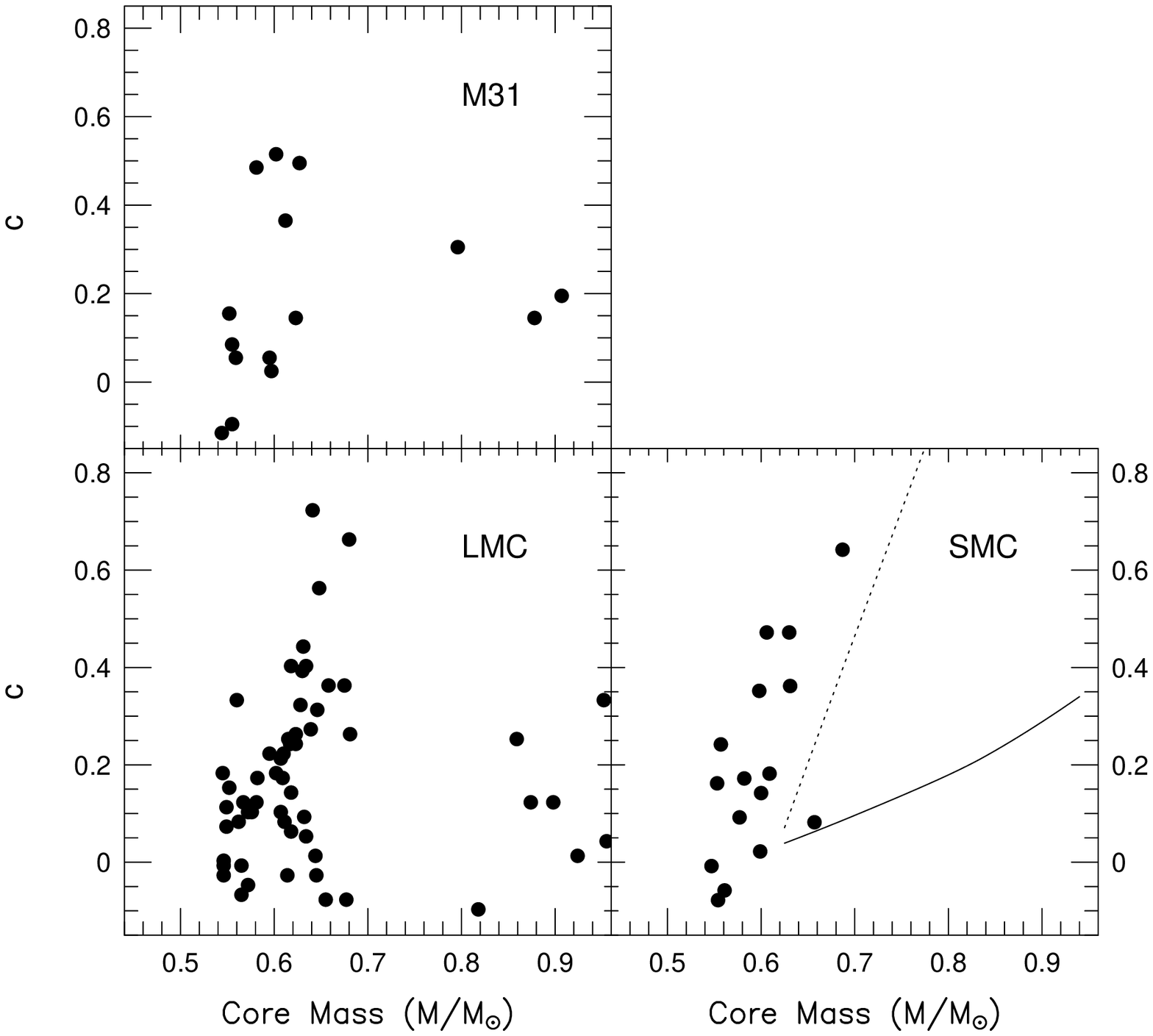}{6.0in}{0.}{80.}{80.}{-240}{0}
\caption{The variation of extinction (computed from measurements of the 
Balmer decrement) vs.~derived core mass for samples of bright
PNe in M31 and the Magellanic Clouds.  Although the scatter is substantial,
all three galaxies show a statistically significant correlation between
circumstellar extinction and core mass.   The solid line in the SMC
panel is our simple extinction model using a nebular
radius equal to that of a median planetary (0.07~pc).  The dashed line 
represents a similar extinction model, but with a radius of only
0.02~pc.  The vertical placement of the curves are somewhat arbitrary, in
that they depend on the assumed AGB mass ejection velocity and the
gas-to-color-extinction ratio.}
\end{figure}

\clearpage
\begin{figure}
\figurenum{2}
\plotfiddle{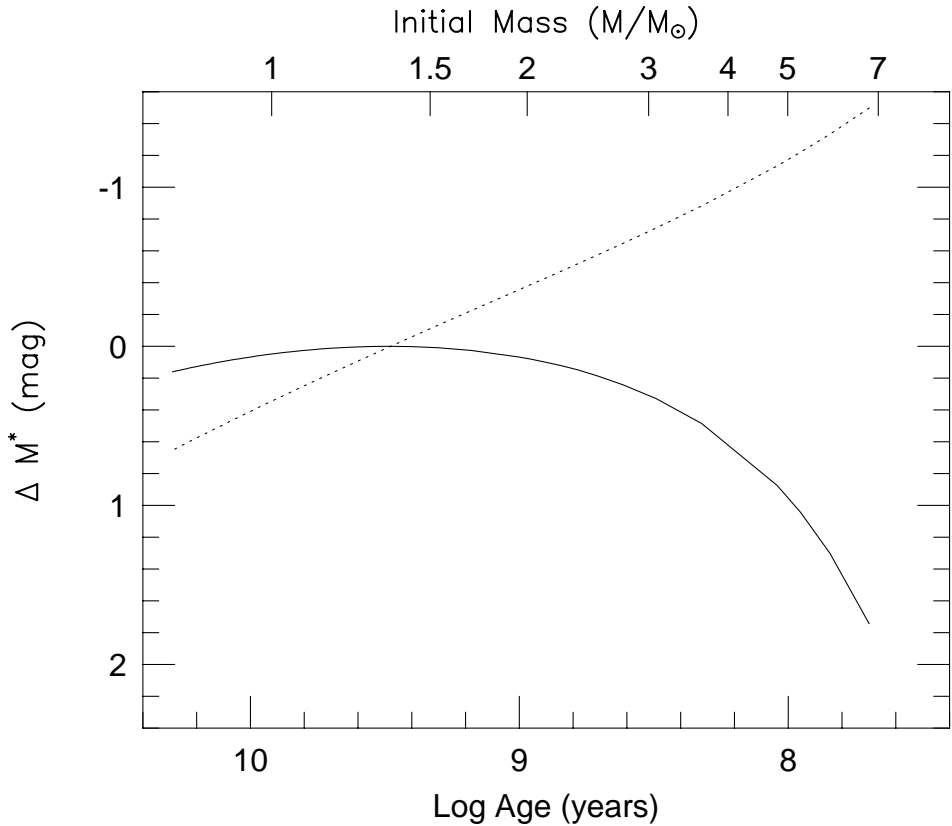}{6.0in}{0.}{80.}{80.}{-240}{-90}
\caption{The change in the PNLF cutoff magnitude ($M^*$) as a function of 
population age.  The dotted line shows the prediction of Jacoby (1996), which 
does not include the effects of circumstellar extinction; the solid line
shows the same calculation, but with a slope of $\sim 6$ for the 
extinction-core mass relation.   Note that when circumstellar extinction
is included, the PNLF cutoff becomes insensitive to age for populations
older than $\sim 1$~Gyr, and that the cutoff for extremely young populations 
is fainter than that for older systems.  Since real galaxies have a mix of 
stellar populations, the older stars will always dominate the bright end 
of the PNLF{}.  This helps to explain why the technique of using bright 
planetary nebulae to measure extragalactic distances works well in both spiral
and elliptical galaxies.}

\end{figure}

\end{document}